\documentclass[iop,apj,tighten,numberedappendix]{emulateapj}
\usepackage{amssymb}
\usepackage{amsmath}
\usepackage{epsfig}
\usepackage{graphicx}
\usepackage{natbib}
\usepackage{color}
\usepackage{enumerate}

\newcommand{\bmath}[1]{\mbox{ \boldmath $\!#1\!$ \unboldmath}}

\journalinfo{The Astrophysical Journal, in press}

\begin{document}

\title{A unified Lense-Thirring precession model for 
optical and X-ray quasi-periodic oscillations in black hole binaries}

\shorttitle{QPOs in black hole binaries}
\shortauthors{Veledina, Poutanen, \& Ingram}

 \author{Alexandra Veledina\altaffilmark{1}, Juri Poutanen\altaffilmark{1}, and
 Adam Ingram\altaffilmark{2}} 
 
\affil{
$^1$Astronomy Division, Department of Physics, P.O. Box 3000, 
FI-90014 University of Oulu, Finland; \\
alexandra.veledina@oulu.fi, juri.poutanen@oulu.fi \\
$^2$Astronomical Institute ``Anton Pannekoek'', 
University of Amsterdam, Postbus 94249, 1098 GE Amsterdam, The Netherlands}
 
 \submitted{Received 2013 June 10; accepted 2013 October 9; published 2013 ??? }

\begin{abstract} 
Recent observations of accreting black holes reveal the presence of  quasi-periodic oscillations
(QPO) in the optical power density spectra.
The corresponding oscillation periods match those found in the X-rays, implying a common origin. 
Among the numerous suggested X-ray QPO mechanisms, some may also work in the optical.
However, their relevance to the broadband -- optical through X-ray -- spectral properties have not
been investigated.
For the first time, we discuss the QPO mechanism in the context of the self-consistent spectral model.
We propose that the QPOs are produced by  Lense-Thirring precession of the hot accretion
flow, whose outer parts radiate in the optical wavelengths.
At the same time, its innermost parts are emitting the X-rays, explaining the observed
connection of QPO periods.
We predict that the X-ray and optical QPOs should be either in phase or shifted by half a period,
depending on the observer position. 
We investigate the QPO harmonic content and find that the variability amplitudes at the fundamental
frequency are larger in the optical, while the X-rays are expected to have strong harmonics.
We then discuss the QPO spectral dependence and compare the expectations to the existing data.
\end{abstract}

\keywords{accretion, accretion disks -- black hole physics -- radiation processes: nonthermal  -- 
X-rays: binaries} 

\section{Introduction}

Accreting black holes (BH) remain among the most fascinating objects
being studied since the very beginning of the X-ray era. 
The observed X-ray radiation is variable on a wide range of timescales, displaying
dramatic changes in spectral shape between the power law dominated
hard state and the quasi-thermal soft state over timescales of weeks
\citep{ZG04}. 
Strong variability is also seen on far shorter (down to $\sim$10~ms) timescales. 
Among the most prominent features commonly observed in the power spectral density (PSD) of BH
binaries are low-frequency quasi-periodic oscillations (QPOs)
\citep{RM06,DGK07}, with a characteristic frequency which evolves from $\sim$0.1--10~Hz at the
spectral transitions from the hard to soft state. 
A growing number of low-mass X-ray binaries have also been found to exhibit similar QPOs in the
optical and UV PSDs \citep{Motch83,Motch85,Imam90,SteiCam97,HHC03,DGS09,GDD10}.
The optical, UV and X-ray QPOs in XTE~J1118+480 have been observed to share a common
(within uncertainties) characteristic frequency whilst evolving over nearly
two months of observations \citep{HHC03}. 
In the X-rays, the QPO frequency is correlated with the low-frequency break of the broadband
noise \citep{WvdK99,BPK02}. Hints of such a correlation in the optical
can also be found in the existing data \citep{GDD10,HHC03},
however the significance is too low to make any conclusive statements. 
It is therefore suggestive that the X-ray and optical QPOs are formed by a common mechanism which
is somehow related to the production of aperiodic variability (broadband noise). 
However, there is no consensus in the literature as to the origin of the X-ray QPO
and so far no optical QPO mechanism has been suggested.

Proposed X-ray QPO mechanisms are generally based either on the
misalignment of the BH and binary system spins \citep[e.g.,][]{SV98} or on
oscillation modes of the accretion flow itself \citep[e.g.,][]{WSO01}.
Many of them, however, discuss the QPO production separately and do
not consider its relation to the aperiodic X-ray variability.
Probably the most promising QPO model to date was proposed in
\citet{IDF09}, where the oscillations arise from the precession
of orbits around the BH due to misalignment of the BH and orbital
spins, known as Lense-Thirring precession. 
Similar models proposed earlier considered the precession of a test mass \citep{SHM06}, leading to a
strong dependence of the predicted QPO frequency on BH spin, inconsistent with the observations. 
The model of \citet{IDF09} considers the precession of the entire hot flow, which leads to a much
weaker spin dependence. 
We note that the precession of the entire flow as a solid body can only be possible in the case of a
hot geometrically thick accretion flow \citep{FB07}, while a cold thin
disk would produce a steady warp in the plane perpendicular to the BH
spin \citep{BP75,KP85}. 
Recently, it was shown that the physical parameters of the hot flow picked to match the observed QPO
frequency are also consistent with those required to produce the characteristic
frequencies of the broad band noise \citep{ID11}, thus a global
connection between the QPO and aperiodic variability was established.
 
This variability model ultimately requires a
geometry where the cold \citep{SS73} accretion disk is truncated at
some radius and the BH vicinity is occupied by some type of the hot
accretion flow (\textit{the truncated disk model}; \citealt{Esin97,PKR97}). The power law
component of the X-ray spectrum can be understood in this geometry as
Compton up-scattering of cold seed photons by hot electrons in the
flow. The seed photons are often assumed to be provided by the disk,
however in the hard state when the truncation radius is large
($\sim$30$R_{\rm S}$ where $R_{\rm S}=2GM/c^2$ is the Schwarzschild radius), the
luminosity of disk photons incident on the flow is insufficient to
reproduce the observed spectra and an additional source of seed
photons is required. This can be provided by synchrotron radiation in
the flow. Compton up-scattering of the resulting internally produced
seed photons (the synchrotron self-Compton mechanism) can successfully
reproduce the observed hard state spectra \citep{MB09,PV09}. Let us note
that an array of accretion geometries can explain the observed spectra
but the Lense-Thirring QPO model is compatible only with the truncated
disk model considered here.

Lately, it was shown that the hot flow synchrotron emission can
contribute to the optical and even IR (OIR) wavelengths \citep{VPV13}, thus
these wavelengths are tightly connected to the X-ray component. Hence,
the spectral model also predicts the connection between the OIR
and the X-ray aperiodic variability. However, simultaneous analysis of
light-curves in these wavelengths \citep{Motch83,Kanbach01,HHC03,DGS08,GDD10}
has revealed a complicated connection between them, which can be
understood if one considers several components contributing to the
optical \citep{VPV11}, namely the hot accretion flow and the
reprocessed radiation.  Additionally, the jet optically thin synchrotron  emission
can also be significant \citep[e.g.][]{CMO10}. These three components
presumably can be responsible for the OIR QPOs.

In Section~\ref{sect:models_setup} we develop a quantitative model for the low-frequency OIR and X-ray QPOs, 
which are produced  by Lense-Thirring precession of a hot accretion flow. 
We make predictions for the OIR QPO profiles and compare them to those expected in the X-rays. 
We then calculate the phase-lags and harmonic content expected from this model. 
In Section~\ref{sec:discussion}, we  discuss the limitations of the model and possible improvements and 
compare the model predictions with the data. We summarize our findings in Section~\ref{sec:summary}.

\section{Hot flow QPO model}
\label{sect:models_setup}

\subsection{Assumptions}
\label{sect:assump}

\begin{figure}
 \plotone{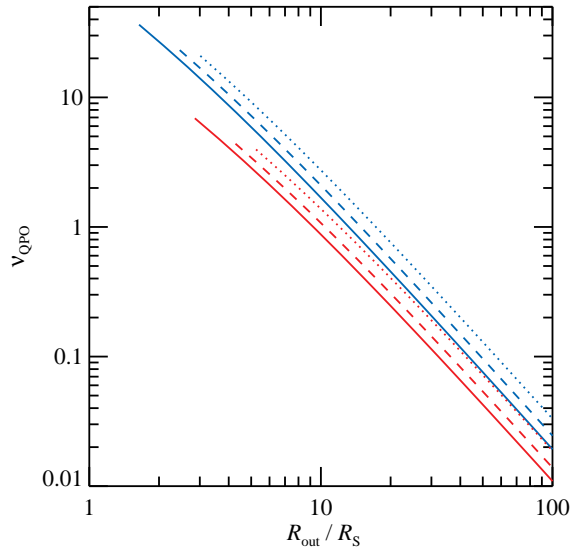}
\caption{
Possible QPO frequencies as function of the hot flow outer radius ($R_{\rm out}$), calculated for a
10$M_\odot$ BH according to equation~(43) of \citet{FB07}, with inner radius equal to 
$R_{\rm in}=1.5\ (H/R)^{-4/5}a^{2/5}R_{\rm S}$ \citep{LOP02}, for spins of $a=0.2$ (solid), 0.3 (dashed) and
0.5 (dotted).
Red lines correspond to $H/R=0.2$ and blue lines correspond to $H/R=0.4$.
} \label{fig:qpo_freq}
\end{figure}

In this section, we outline our assumptions. 
We consider a hot accretion flow extending from some inner radius to the truncation radius of the
cold outer disk and a spin axis misaligned with that of the BH by some angle $\beta$. 
In the Kerr metric, test particle orbits out of the plane of BH rotation undergo Lense--Thirring
precession due to the dragging of inertial frames, with a frequency 
$\nu_{\rm LT}(R) \propto \sim R^{-3}$. 
In the case of a hot accretion flow, this differential precession creates a warp which propagates
through the flow on a time scale shorter than the precession period. 
This can cause the entire flow to precess as a solid body at a
frequency given by a surface density weighted average of $\nu_{\rm LT}(R)$ \citep{FB07}. 
If the flow is assumed to extend from the disk truncation radius to the
inner most stable circular orbit, the dependence of the precession frequency on BH spin is
too strong and the maximum precession frequency is too high to be consistent with observations.
In order to put the Lense--Thirring frequencies into the observed QPO range, from $\sim$0.03~Hz
\citep{VCG94a} up to $\sim$13~Hz \citep{Rem99}, one needs to account for the torque created by
frame dragging, which truncates the inner flow at a fairly large (inner) radius
($\gtrsim3R_{\rm S}$, see equation~22 of \citealt{LOP02}, where $x=1$ should be taken) 
and effectively cancels out much of the spin dependence of precession \citep{IDF09}.
This effect has since been seen in simulations \citep{Fragile09}.
We reproduce the dependence of the QPO frequency on the outer radius of the hot flow in
Figure~\ref{fig:qpo_freq}.
Since we aim to model the hard state, we assume a fairly large truncation radius of 
$R_{\rm out}=30R_{\rm S}$,
leading to a predicted precession frequency consistent with the observed QPO frequency in this
state ($\lesssim0.3$~Hz).
When $R_{\rm out}$ decreases (corresponding to transition to the
hard-intermediate state), $\nu_{\rm QPO}$ grows up to 5--20 Hz.

Although the condition on the flow height-to-radius ratio, $H/R$, for the fast propagation of
warps which allows solid body precession is $H/R >\alpha\sim0.1$,
where $\alpha$ is the dimensionless viscosity parameter \citep{PP83,FB07}, $H/R$ significantly
less than unity is likely \citep{KFM08}.
Thus, we approximate the flow as a flat disk. 
This approximation is satisfactory for systems with a low enough inclination for the flow never to
be seen edge-on at any point in the precession cycle. 
We illustrate in our plots the regions of parameter space where this condition is not met. 
In order to calculate the emitted flux, we must make an assumption about the radial dependence
of energy dissipation per unit area in the flow. 
We assume the standard profile for a thin disk \citep{SS73}
\begin{equation}\label{eq:SS73_profile}
  q(R) \propto R ^{-3} \ \sqrt{1-\frac{R_{\rm in}}{R}}  ,
\end{equation}
where $R_{\rm in}$ is the hot flow inner radius, which we assume to be equal 3$R_{\rm S}$
(for parameters in Figure~\ref{fig:qpo_freq} it takes the values $\sim$2--5$R_{\rm S}$).
The profile is not strictly appropriate for a hot flow in general relativity \citep{NT73,KFM08},
but is fine for illustration in the absence of a standard equivalent for hot flows. 
The details of the energy dissipation profile do not play an important role. 
For calculations of the angular dependence of the emergent radiation 
we take into account relativistic effects using the Schwarzschild metric (see Sect.~\ref{sect:profiles}). 
Since we are considering Lense-Thirring precession, this is obviously only an approximation to the
more appropriate Kerr metric. 
However, the Schwarzschild metric provides a very good approximation at distances more than
$3R_{\rm S}$ from the BH.

We take into account only direct images, because the higher-order images are  much weaker. 
The reason is that  the hot flow we consider has optical depth of the order of unity, thus most of the
photons initially emitted away from the observer would be blocked by the flow itself (or by the outer cold disk). 
The only possibility for those photons to reach the observer is to go through the
rather small gap between the inner edge of the hot flow ($\sim3R_{\rm S}$) and the black hole.
In addition, photons coming in the higher-order images are generally emitted at grazing angles to the 
flow surface and therefore their strength is diminished by the corresponding cosine factor.  
In order to estimate the accuracy of the described approach, we compared fluxes 
obtained using our light tracing procedure with those obtained using the {\sc geokerr} code 
in Kerr geometry (\citealt{DA09}; Adam Ingram et al., in prep.) including all higher-order images,
but accounting for obscuration by the hot flow and the outer cold disk.  
For the parameters considered in this work, we find our results to be accurate within a few per
cent for any spin value up to $a=0.998$, and for more realistic spins of $a=0.3, 0.5$ the difference is below 1\%. 

\subsection{Geometry}
\label{sect:geom}

\begin{figure}
 \plotone{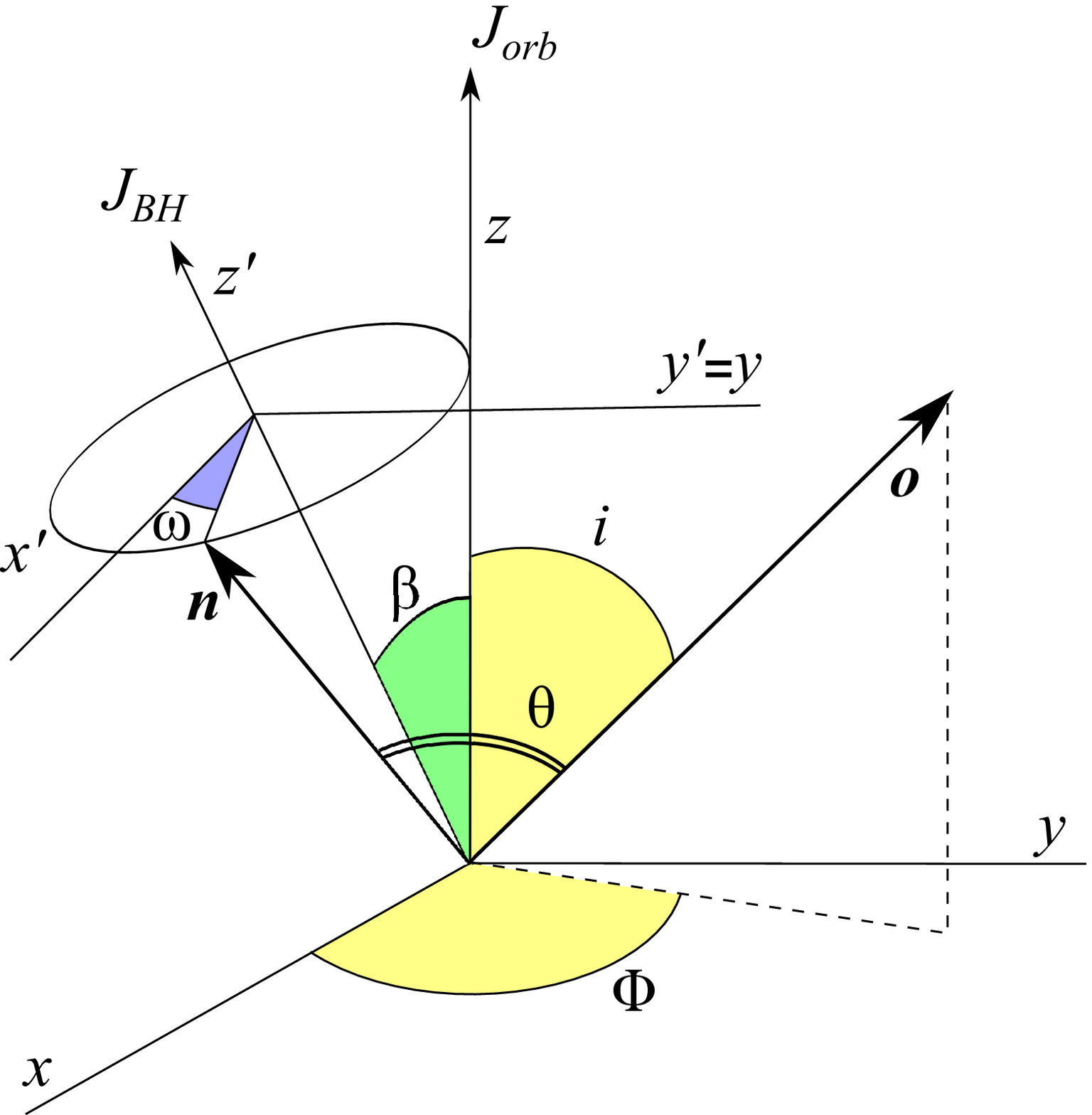}
\caption{ 
Schematic representation of the hot flow producing X-ray and optical QPOs.
Coordinate systems connected with the orbital plane $xyz$ and with the BH spin $x'y'z'$ are shown.
Plane $xy$ coincides with the orbital plane and $y$ is parallel to $y'$.
Axis $z'$ is aligned with the BH spin, which is inclined by the angle $\beta$ to the orbital spin.
Position of the observer $\hat{\bmath{o}}$ is described by the azimuthal angle $\Phi$ and binary
inclination $i$.
The current position of the hot flow normal $\hat{\bmath{n}}$ is characterized by
the precession angle(phase) $\omega$.
Relative to the direction to the observer, it makes an angle $\theta$,  which depends on $\omega$.
} \label{fig:geometry_flow}
\end{figure}

Figure~\ref{fig:geometry_flow} shows the geometry we consider for
the precessing hot accretion flow. We define the coordinate system
$xyz$ tied to the binary orbital plane and the system $x'y'z'$,
associated with the BH spin. The $z$-axis is perpendicular to the
binary orbital plane, $z'$ is aligned with the BH spin and $\beta$ is
the angle between them. The $x$-axis is defined by the intersection of the
orbital plane and the plane formed by the $z$ and $z'$ axes (i.e. the
projection of the $z'$-axis onto the $y$-axis is always
zero). Finally, the $x'$-axis also belongs to the plane formed by
vectors $z$ and $z'$, and forms angle $\beta$ with the $x$-axis
(i.e. $y'=y$). The position of the observer is described by the vector
$\hat{\bmath{o}}$ which, in $xyz$ coordinates, is given by
\begin{equation}\label{eq:obs_vec}
\hat{\bmath{o}}=(\sin i \cos\Phi, \sin i \sin\Phi, \cos i), 
\end{equation}
where
$i$ is the binary inclination and $\Phi$ is the azimuth of the
observer measured from the $x$-axis. The instantaneous normal to the
hot flow is denoted by $\hat{\bmath{n}}$. It precesses around the BH
spin axis ($z'$) with the precession angle $\omega$ measured from the
$x'$-axis. Thus $\hat{\bmath{n}}$ aligns with the $z$-axis when
$\omega=\pi$ and has a maximal misalignment of $2\beta$ when
$\omega=0$. It is given, in $x'y'z'$ coordinates, by
$\hat{\bmath{n}}=(\sin\beta  \cos\omega , \sin\beta\sin\omega,\cos\beta)$. Note, we use a hat
to denote unit vectors throughout.

We are interested in the orientation of the hot flow surface relative to
the observer, namely the scalar product $\hat{\bmath{o}} \cdot
\hat{\bmath{n}} = \cos\theta$. In order to calculate it, we write
$\hat{\bmath{n}}$ in $xyz$ coordinates by rotating the $x'y'z'$
coordinate system counter-clockwise by angle $\beta$ around the
$y'$-axis to get 
\begin{equation}\label{eq:disc_normal}
\hat{\bmath{n}} = (\sin\beta \cos\beta(1+\cos\omega),  \sin\beta \sin\omega,
\cos^2\beta -\sin^2\beta\cos\omega). 
\end{equation}
Thus, the equation for $\cos\theta$ is
\begin{eqnarray}\label{eq:cos_theta}
   \cos\theta &=& \sin\beta \cos\beta \sin i  \cos\Phi\  (1+\cos\omega)  \\ 
           &+& \sin\beta \sin\omega \sin i  \sin\Phi + \cos  i\ (\cos^2\beta -\sin^2\beta\cos\omega). \nonumber
\end{eqnarray}

\subsection{Formalism} 
\label{sect:profiles}

We consider a simplified problem with a precessing flat disk (slab). 
The specific flux observed far from the BH  from a ring of 
thickness $dR$ at radius $R$ in the direction that makes an angle $\theta$ to the surface normal  
can be expressed as
\begin{equation}\label{eq:flux_azimuth_ave2}
 dF_E (R, \theta) = \frac{R\ dR}{D^2}  q_E(R) f (R,\theta).
 \end{equation}
Here, $D$ is the distance to the observer, $q_E(R)$ is the surface
flux per energy interval at a given radius and the factor
$f(R,\theta)$ accounts for the angular dependence of the observed flux. 
The latter factor depends on the spectral slope of the radiation. 

The flux observed from the whole hot flow is then 
\begin{equation}\label{eq:flux_azimuth_ave3}
 F_E (\theta) = \frac{ \int q_E(R)\ R\ dR}{D^2} \  \overline{f}(\theta), 
 \end{equation}
where we introduced  the radially averaged angular factor:
\begin{equation}\label{eq:flux_ave_radii}
 \overline{f}(\theta) = \frac{ \int q_E(R)\ R\ dR \ f (R,\theta) } { \int q_E(R)\ R\ dR }.
 \end{equation}
 
In general, the specific intensity emerging from a surface element depends on the
zenith angle $\zeta'$ (primed  are quantities in the comoving frame). 
We consider here two cases for the emission pattern from the flow surface:
\begin{enumerate}
  \item X-ray emission from accreting BHs is produced by
    Comptonization in an optically translucent flow with Thomson
    optical depth $\tau\sim1$. 
In this case, the angular dependence of
    the specific intensity can approximately be described by
    \citep{PG03}
        \begin{equation}\label{eq:compton_pattern}
          I'(\zeta') \propto 1 + b \cos\zeta' ,
        \end{equation}
        where $b\approx -0.7$ (for exact solutions see
        \citealt{ST85,VP04}). 
  \item Optical synchrotron radiation is produced in the outer parts of the flow, in a
    partially self-absorbed regime. The dominant contribution to
    the observed flux comes from those parts, which have an optical depth for synchrotron
    self-absorption $\tau_{\rm SA}\sim 1$. Assuming a tangled
    magnetic field and homogeneous source function throughout the
    vertical extent of the flow, the intensity of radiation can be
    expressed as
        \begin{equation}\label{eq:synchro_pattern}
          I'(\zeta') \propto 1 - \exp\left( -\tau_{\rm SA}/\cos \zeta' \right).
        \end{equation}
\end{enumerate}

\begin{figure}
\plotone{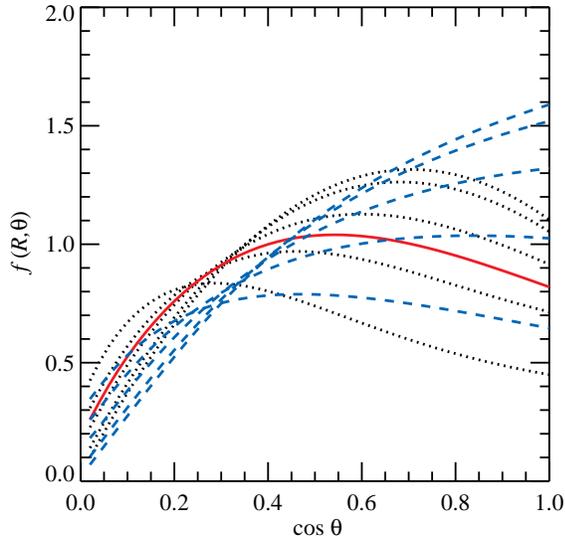}
\caption{
Angular dependence of the observed flux $f(R,\theta)$ as a function of the angle between the observer direction and the
normal to the flow.
The black dotted lines represent result for the X-ray emission pattern (equation~\ref{eq:compton_pattern}) corresponding to Comptonization
in an optically translucent flow.
The blue dashed lines correspond to the partially self-absorbed synchrotron emission expected in the
optical (given by the pattern from equation~\ref{eq:synchro_pattern}). 
Curves from the top to the bottom correspond to $R/R_{\rm S}=100, 30, 10, 5$ and $3$. 
The solid red line corresponds to the flux $\overline{f}(\theta)$ 
from the accretion flow between 3 and 30$R_{\rm S}$ emitting the X-rays with the standard emissivity profile (\ref{eq:SS73_profile}).
} \label{fig:flux_theta}
\end{figure}

We assume that the emergent spectrum from all surface elements is a power
law with  photon index $\Gamma=1.7$ for the X-rays and $\Gamma=1$ for the optical emission. 
This allows us to   approximate the energy dissipation profile as $q_E(R)\propto q(R)$. 
To compute the factor $f(R,\theta)$ we take into account gravitational redshift, Doppler effect, 
relativistic aberration, time dilation and light bending in the Schwarzschild
metric following techniques presented by \citet{PG03} and \citet{PB06}. 
Figure~\ref{fig:flux_theta} shows the resulting angular flux
dependence with  X-rays and optical represented  by black
dotted and blue dashed lines, respectively. 
We plot the function $f(R,\theta)$ for narrow
rings at $R/R_{\rm S}=$ 100, 30, 10, 5 and 3 from top to bottom.
We also show (by red solid line) the angular distribution of the flux from the accretion flow 
spread from 3 to 30$R_{\rm S}$, $\overline{f}(\theta)$, emitting the X-rays according to the 
standard profile given by equation~(\ref{eq:SS73_profile}). 
At large radii $R\gtrsim30R_{\rm S}$, the relativistic effects are negligible 
resulting in the angular dependence $f(R,\theta)\propto  I'(\theta) \cos\theta$.

\begin{figure*}
\centerline{\epsfig{file=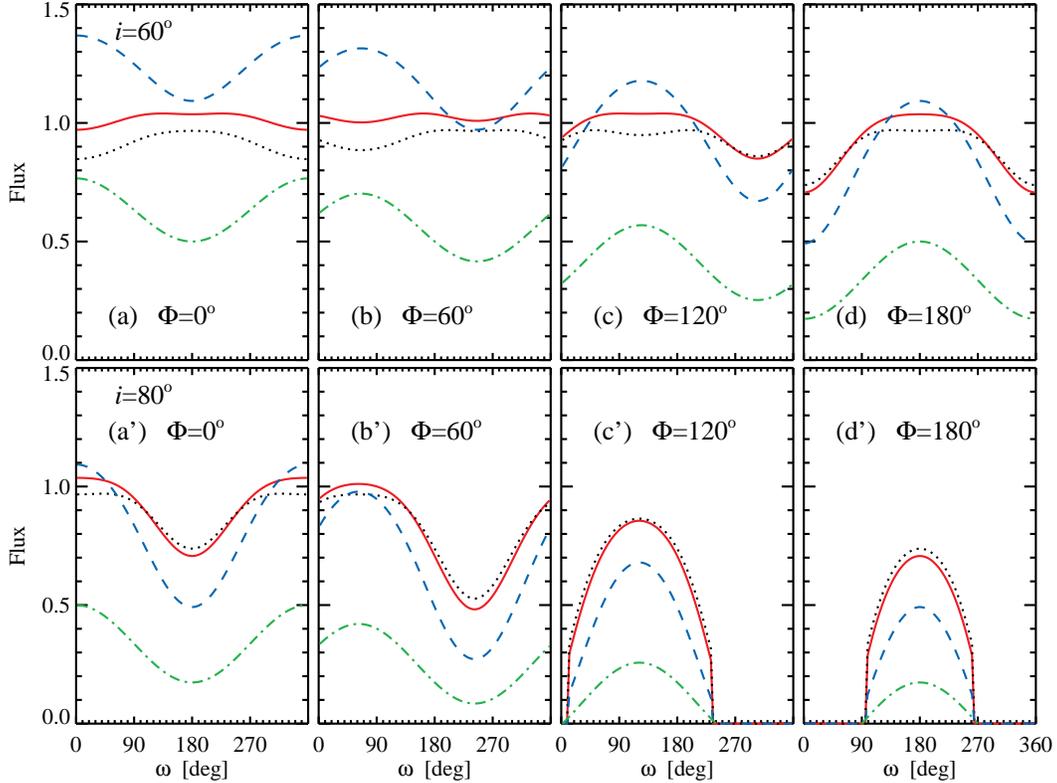,width=14.0cm}}
\caption{
Possible QPO waveforms for the four cases of the observer's azimuthal angle.  
The parameters are: $\beta=10\degr$, 
$i=60\degr$ (upper panels) and $i=80\degr$ (lower panels).
X-ray profiles are calculated 
for a ring at $R=5R_{\rm S}$ (black dotted  lines) and for the full hot flow 
extending from $3$ to $30R_{\rm S}$ (red solid  lines). 
The optical profiles are for $R=30 R_{\rm S}$ (blue dashed lines).
Green dash-dotted lines represent variations of $\cos\theta$ for each case.
} \label{fig:profiles}
\end{figure*}

\begin{figure*}
\centerline{\epsfig{file=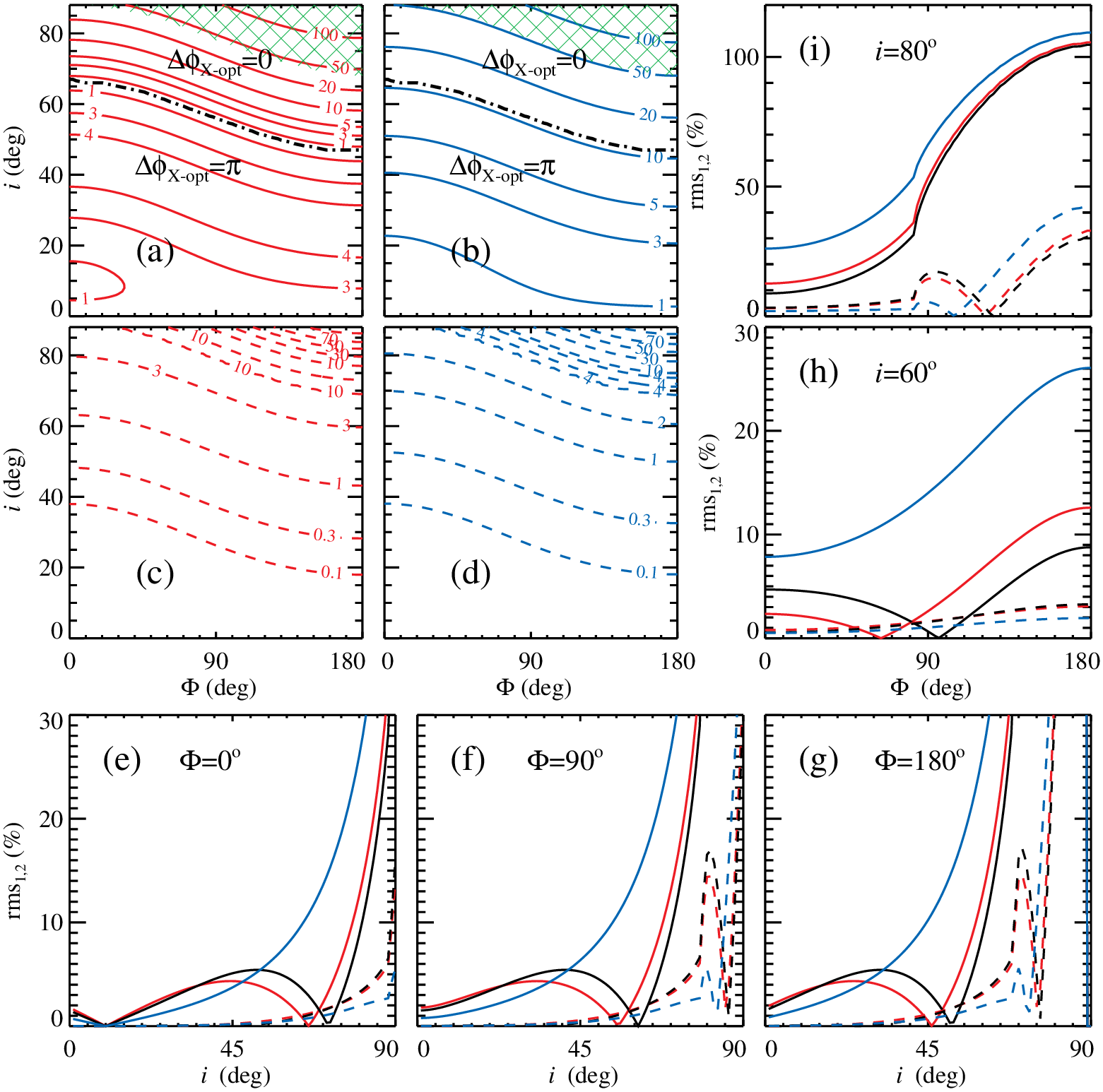,width=14.0cm}}
\caption{
Fractional  RMS amplitudes (in percent) as functions of $i$ and $\Phi$.
The left-upper four panels present the 
contour plots for X-ray (a,c) and optical (b,d) fractional  RMS amplitudes on the plane $i$--$\Phi$. 
The upper panels  (a and b) are for the fundamental, while the two lower panels (c and d) show the harmonic. 
Eclipses are possible in the shaded regions marked on the top panels. 
The dot-dashed black line separates the regions, 
where the phase difference between the optical/X-ray fundamental
QPOs shifts from 0 to $\pi$.
The amplitude of the X-ray fundamental turns zero along this line.
Three bottom panels present the cross-section of these plots 
at the azimuthal angles $\Phi=0\degr$ (panel e), $90\degr$ (panel f), and $180\degr$ (panel g).
The panels on the right show 
the cross-section  at two inclinations $i=60\degr$ (panel h) and $80\degr$ (panel i).
The X-rays (red curves) are assumed to originate from a hot disk extending from $3$ to $30R_{\rm S}$
emitting according to the standard emissivity profile given by equation (\ref{eq:SS73_profile}), 
while the optical (blue curves) corresponds to a narrow ring at $R=30R_{\rm S}$. 
The solid curves present the RMS for the fundamental and 
the dashed curves give the harmonic. 
The  RMS for the X-rays from a narrow ring at $R=5R_{\rm S}$  
are shown in  the  panels (e--i) by black  curves,
which show similar  behavior to the full hot flow case. 
} \label{fig:rms}
\end{figure*}

\subsection{Results}

As a first step, we take the
X-ray and optical emission to originate from narrow rings with
$R=5R_{\rm S}$ and $R=30R_{\rm S}$, respectively, 
and compute corresponding QPO waveforms.  
We set $\beta=10\degr$ and $i=60\degr,\,80\degr$. 
The profiles $f(R,\theta)$ (with $\theta$ given by equation~(\ref{eq:cos_theta})) 
are shown  in Figure~\ref{fig:profiles} for four different observer
azimuths: $\Phi=0\degr$, $60\degr$, $120\degr$ and $180\degr$. 
Again, the black dotted lines are X-rays and the blue dashed line represents the optical. 
Clearly the phase difference between optical and X-ray
profiles  depends on the observer's azimuth and inclination: 
for $i=60\degr$ and large $\Phi$ the two light-curves are in phase, while for low $\Phi$ they are
$180\degr$ out of phase. 
This is because the range of $\cos\theta$ covered
as the precession angle $\omega$ unwinds from $0$ to $2\pi$ depends on
the chosen set of parameters ($\beta$, $i$, $\Phi$). 
The range of $\cos\theta$ for every case is shown in Figure~\ref{fig:profiles} with
a green dash-dotted line. For small $\cos\theta$ (panels c and d), both
optical and X-ray observed fluxes are increasing functions of
$\cos\theta$ (see Figure~\ref{fig:flux_theta}), while for large
$\cos\theta$ (panels a and b) optical flux is a growing and X-ray flux
is a decreasing function of $\cos\theta$. The later is reflected in
the anti-correlation of the two light-curves.

For inclinations $i\gtrsim80\degr$, eclipses are possible when the observer is situated at
the opposite side from the BH spin projection on the orbital plane, i.e. at $\Phi>90\degr$. 
In such a situation, the harmonic content is dramatically increased and 
the X-ray and optical light curves are always in phase, 
as here $\cos\theta<0.5$ (thus both fluxes have monotonic dependence on $\cos\theta$, see
Figure~\ref{fig:flux_theta}). 
However, we need to keep in mind that at such large inclinations the central source might be
blocked completely by the outer rim of the cold disk. 

We note that the amplitude of the optical QPO for every $\Phi$ is
higher than in the X-rays. Again, the behavior can be understood from
Figure~\ref{fig:flux_theta}: the dependence of the optical flux (blue
dashed line at $30R_{\rm S}$) on $\cos\theta$ is much stronger than
that in the X-rays (black dotted line at $5R_{\rm S}$). 
This occurs both due to the different emission pattern and because 
the relativistic effects are stronger at smaller $R$
(due to stronger gravitational curvature and faster Keplerian motion). 
 
For the cases (b)--(d), the X-ray flux maximum is reached not when 
$\cos\theta$ is maximal, i.e.  when the flow is mostly face-on, 
but when it is inclined at a larger angle (see Figure~\ref{fig:flux_theta}).
This leads to the double-peak structures in the X-ray profiles, unlike
in the optical (optical flux reaches maximum only at maximal
$\cos\theta$, i.e. when the flow is mostly face-on). This gives the
X-ray QPO profiles a stronger harmonic content than the optical. 

The X-ray light curves expected from the whole 
hot flow occupying the region 3--30$R_{\rm S}$ 
can be computed from the factor  
$\overline{f}(\theta)$ (with $\theta$ given by equation~(\ref{eq:cos_theta})). 
They are shown by red solid  lines in Figure~\ref{fig:profiles}. 
We see that their behavior is very similar to that corresponding to a narrow ring at $5R_{\rm S}$. 
This is due to the similarities in the angular dependence of the fluxes (see
Figure~\ref{fig:flux_theta}, solid line and dotted line corresponding to $5R_{\rm S}$).  
The emission pattern from a ring at  $\sim7R_{\rm S}$ would be  nearly identical to that of the whole
extended flow, and so assuming all of the flux to come from close to this radius provides a good
approximation.

We analyze the harmonic content of our light curves by directly computing Fourier coefficients. 
We show the fractional root-mean-square (RMS) variability amplitudes of the first harmonic
(referred to below as the fundamental) and the second harmonic (hereafter referred to as the
harmonic) as the contour plots at the plane observer azimuth $\Phi$ --  observer inclination $i$
(Figure~\ref{fig:rms}).
The azimuthal angle spans the interval from 0 to 180\degr, because the results 
depend on $\cos\Phi$ only.  
The X-rays are assumed to be coming from the whole flow extending from $3$ to
$30R_{\rm S}$ with a standard emissivity profile given by equation (\ref{eq:SS73_profile}).
The X-ray RMS is shown by red lines. 
The optical again is assumed coming from $30R_{\rm S}$ and its RMS is shown by blue lines.

Behavior of the X-ray and optical fundamental is shown in Figures~\ref{fig:rms}(a) and (b),
respectively.
The optical RMS increases monotonically with both $i$ and $\Phi$. 
The X-ray RMS drops to zero at inclinations between 45\degr and 65\degr, depending on $\Phi$, where
the light curve shows a double bump structure (thus, the harmonic here is non-zero).
At higher inclinations RMS increases again. 
At high inclinations and large $\Phi$ in the shaded region, 
self-eclipses of the disk become possible, causing dramatic increase of RMS.
The dot-dashed black line separates the regions, where the phase difference between the optical/X-ray fundamental
QPOs shifts from 0 to $\pi$.
Also the amplitude of the X-ray fundamental turns zero along this line.

Figures~\ref{fig:rms}(c) and (d) present the contour plot of the harmonic RMS for X-rays and
optical, respectively. 
Here the RMS is monotonically growing with both $i$ and $\Phi$
up to rather large $i\gtrsim70\degr$ (i.e. $\sim 90\degr-2\beta$). 
At large inclinations  eclipses become possible
(see right panels of Figure~\ref{fig:profiles}). 
In that case,  both RMS show very erratic behavior and depend strongly on $i$,
first rapidly rising and then dropping to  
nearly  zero (at somewhat different places of the plane shown by black dashed lines), but here
the  higher harmonics are strong. This is easily seen in Figures~\ref{fig:rms}(e)--(g) 
which show the RMS behavior as a function of $i$ for fixed $\Phi$.

We see that, for the chosen parameters, the optical fundamental is stronger than the X-ray fundamental for all observer
azimuths whereas the X-ray harmonic is nearly always stronger than the
optical harmonic.
Without eclipses, the optical harmonic is typically ten times weaker than the fundamental,
while the X-ray harmonic can be even stronger than the corresponding fundamental. 
For better vizualization we also plot a cross-section of these contour plots  
in Figures~\ref{fig:rms}(h) and (i) for two representative inclinations $i=60\degr$ and $80\degr$,
respectively. 
The first is the mean possible inclination of an arbitrary observer, 
while the second is chosen to show the role of eclipses. 
Here we additionally show the case of the X-rays from a narrow ring at $5R_{\rm S}$ by black
lines. 
In panel (h), we clearly see a drop of the fundamental X-ray RMS at $\Phi=50$--100\degr 
and in panel (i) a rapid grow of RMS when eclipses become possible at $\Phi\sim 80$\degr.

\section{Discussion}
\label{sec:discussion}

 \subsection{Comparison with observations} 
 
Our results show that the X-ray light-curves have strong harmonics, while optical harmonics are
weaker for most of sets ($i$, $\Phi$).
Indeed,  harmonics are often found in the X-ray PSDs \citep[e.g.][]{RCK04}, 
but no optical harmonics have been reported so far.
This can also be seen in the (quasi-) simultaneous data of \citet[][fig.~2]{Motch83}, where both
optical and X-ray PSDs show a QPO at $\sim0.05$~Hz, two harmonics $\sim0.1$~Hz and $\sim0.15$~Hz are
seen in the X-ray, but not in the optical PSD.
Our simulations suggest that the RMS amplitudes of the fundamental harmonic in the optical are  expected to be larger.
Unfortunately, the data  with both X-ray and optical QPOs detected are not strictly
simultaneous, thus the PSDs are plotted in  arbitrary units \citep{Motch83,HHC03}.
In the available simultaneous data of Swift J1753.5--0127, the optical QPO is present with RMS=4--11\%, 
while the upper limit on the X-ray QPO RMS at the same frequency is 3\%  \citep{DGS09}. 
In GX 339--4  situation is similar, the optical QPO is present at RMS=3\%, 
while the upper limit on the X-ray QPO RMS at the same frequency is 7.5\%  \citep{GDD10}.
This is consistent  with our model which predicts that the X-ray QPO is there, just at a low RMS. 
On the other hand, we, of course, cannot rule out there being no QPO at all. 
In order to check the presence of the X-ray QPO in the data (which is related to the optical QPO),
one may search for modulations at the corresponding frequency in the cross-correlation function.
These oscillations are indeed present in the 2007 data on Swift~J1753.5--0127 
\citep[see][Veledina et al. in prep.]{DGS08} and hints of them can also be seen in the
GX~339--4 data \citep[fig.~21]{GDD10}.

In Figure~\ref{fig:qpo_freq} we reproduced the range of possible QPO frequencies (the same for
optical and X-rays) as a function of the hot flow outer radius.
Since the dependence on parameters is not too strong, it is feasible to determine $R_{\rm out}$
for a given $\nu_{\rm QPO}$.
In realistic situation, $R_{\rm out}$ fluctuates together with the mass accretion rate, leading to
a change of the QPO frequency and broadening of the QPO features in the PSD.
We note that the given frequencies were computed under the assumption of the rotation of the hot
flow as a solid body, which breaks down for a large hot flow, limiting the outer radii, where the
QPOs can be observed. 

The lowest QPO frequencies $\sim0.05$~Hz (optical) and $\sim$0.03~Hz (X-rays) were observed in the
hard state.
According to Figure~\ref{fig:qpo_freq}, we find the hot flow outer radius can be
$\sim$100~$R_{\rm S}$, and even larger for $H/R>$0.4 or spin $a>$0.5.
In this case, the optical and IR wavelengths are dominated by the radiation of the translucent
parts of the hot flow (see e.g., \citealt{VPV13}, fig.~3: optical wavelengths are dominated by
the radiation coming from $\sim$30~$R_{\rm S}$).

During the state transition, with the rise of the mass accretion rate, 
the outer shells of the hot flow are gradually replaced by the cold accretion disk 
\citep{Esin97,Esin98,PKR97}.
As the outer radius of the hot flow decreases,  an increase of the QPO frequency occurs.
This is seen in the data as correlated changes in the spectral slope and QPO frequency 
\citep[see e.g.][]{Gilfanov10},
and expected in the precession model of \citet{FB07} and \citet{IDF09}. 
Simultaneously, a sharp drop of the IR radiation is expected, with fluxes at longer wavelengths
dropping just before those at shorter wavelengths.
While the cold accretion disk is sufficiently far away, the X-ray spectrum does not change
significantly, as the amount of cold disk photons are not enough to make it softer.
The noticeable X-ray spectral transition starts when the cold disk is rather close to the BH, at
radii about 10~$R_{\rm S}$ \citep[fig.~5 of][]{VPV13}.
For such a small hot flow, IR wavelengths belong to the optically thick (self-absorbed) part of the
spectrum, which is very hard, with spectral index up to $\alpha=5/2$ ($F_{\nu}\propto\nu^{\alpha}$).
Thus, in the hard-intermediate state we expect to see the QPO feature with frequencies
$\gtrsim$1~Hz in the near-IR, but not in the mid-IR or longer wavelengths, which would likely be
very faint, probably, too faint to be detectable.
The optical/X-ray phase lags in both translucent and opaque cases are
expected to be the same (see below), either 0 or $\pi$ depending on the observer azimuthal angle and inclination.
Maximum optical/near-IR QPO frequency in our model is the same as in the X-rays $\sim$10~Hz (for
a 10$M_\odot$ BH).

We discussed in detail  the OIR QPOs produced only by the hot accretion flow, however these
presumably may originate from other components, such as reprocessed X-ray radiation (for low QPO
frequencies) and the jet (if it is entirely driven by the accretion flow).
It is likely that the component giving major contribution to the
observed flux also produces the QPOs, because of their rather large RMS. 
Simultaneous data of SWIFT~J1753.5--0127 at late stages of the outburst (\citealt{DGS09}, fig.~6,
see also \citealt{CDSG10}, fig.~5) suggest that the optical lies on the continuation of the X-ray
power-law, as expected in the hot flow scenario \citep{VPV13}.
Some irradiation may also play  a role (as supported by the cross-correlation studies of
\citealt{HBM09} during the outburst peak), but the jet is very faint in this object.
Thus the optical QPOs are likely produced by the hot flow.
It is therefore suggestive that the hot flow scenario also works in other objects displaying QPOs,
such as XTE~J1118+480 and GX~339--4.

\subsection{Unaccounted effects}

Because we aimed to have as few parameters as possible, the model we consider is rather simplified.
In a more realistic problem, a number of additional effects  could also be accounted for, namely: 
 the wavelength dependence of the optical profiles, 
  occultations of the hot flow by the cold accretion disk, 
  effects of the hot flow geometry, e.g. its finite and radius-dependent thickness,  
  presence of the large-scale magnetic field, 
  presence of other spectral components in the optical, such as the reprocessed emission and the jet. 
Let us now consider possible consequences of these effects.

\begin{enumerate}
\item The optical QPO profiles were calculated assuming these wavelengths belong to the partially
self-absorbed regime.
However, if the flow is small (e.g., $R\lesssim10R_{\rm S}$), the optical (and even more certainly the IR) 
is likely to be in the self-absorbed part of the spectrum  ($\tau_{\rm SA}\gg1$). 
In this case,  the  intensity of escaping radiation is isotropic and the flux 
depends only on the surface area of the flow projected  to the line of sight modified by relativistic effects. 
The QPO waveforms in this case are not much different from the previously considered case 
of emission at the self-absorption frequency, but the 
RMS of the fundamental is somewhat larger and the harmonic is weaker.

\item
For calculations of the X-ray QPO profiles, we adopted a simple prescription 
for the angular dependence of the specific intensity from the hot flow given by  equation (\ref{eq:compton_pattern}). 
Obviously, it is only an approximation to the real angular dependence that should be computed  
using the full Compton scattering kernel \citep[see e.g.][]{PS96}. 
The intensity $I(\zeta')$ in reality should depend on the photon energy  as well as on $\tau$ and  temperature, 
with both being functions of the accretion rate and the distance from the BH. 
We do not expect that the general topology of the solution will change, but the 
details (e.g. the position of the line separating the phase lag $\Delta\phi_{\rm X-opt}= 0$ and $\pi$ 
in Figure~\ref{fig:rms}) might differ. The result will also depend on the X-ray energy band.

\item
Hard state BHBs are known to have radio jets \citep{Fender06}. 
Theoretical models for the jet launching require
the presence of the large scale magnetic field. 
In that case,  the synchrotron emissivity within the hot flow is no longer isotropic, 
but depends on the pitch angle $\alpha$ to the magnetic field as $\sin^2\alpha$. 
This would lead to a different angular dependence of the escaping intensity. 
For a poloidal field, radiation would be more  beamed along the hot flow surface 
leading to a luminosity angular dependence similar to that in the X-rays. 
In this case, the optical waveforms are expected to have more harmonic content.

\item Possible occultations of the hot flow by coverage of the inner parts of the cold accretion
disk or by its flared outer parts may occur.
These would lead to the asymmetry of the optical (and X-ray) QPO profiles and 
absence of the secondary peak in the X-ray profiles.
Similarly to the considered occultations at high inclinations, this would lead to an 
increase of the fundamental RMS and harmonic content.
Additionally,  the optical and X-ray profiles will not be anymore in phase (or in anti phase).

\item We considered a simple case with a flat precessing disk, but in reality, the hot flow has
finite thickness. This does not have much effect on the profiles for low inclinations. 
Self-eclipses of the hot flow will appear at somewhat lower inclinations, but, on the 
other hand,   emission from the outer side of the hot flow will be 
visible and the eclipses will not be so deep. Thus, the increase in RMS will not 
be so significant and the harmonic content will be somewhat lower.

\item 
When estimating the RMS we assumed that the optical emission is produced 
entirely by the hot flow. In reality, additionally  there is 
reprocessed emission from the outer disk that might be dominant 
at longer  wavelengths   as well as possibly optically thin, soft emission from 
the radio jet  that might contribute to the emission at  shorter wavelengths. 
These components not only contribute to the flux diluting the RMS from the hot flow,  
but also in principle can be variable at the QPO frequency. 
As reprocessing occurs in the outer part of the accretion disk, 
it acts as a low-pass filter \citep[e.g.][]{VPV11},  reducing signal at high frequencies. 
Thus the  reprocessing is not expected to vary at relatively high QPO frequencies.   
Whether the jet can produce QPOs at the same frequency as the inner hot flow 
precesses is not clear. 
The Lense--Thirring effect  generally cannot cause the jet precession unless 
the jet is driven entirely by  the accretion flow    \citep{NK13}. 
Hence, the presence of several components in the spectrum typically 
will reduce the RMS amplitude of the optical 
QPO by a factor $1+r$, where $r=(F_{E, \rm jet}+F_{E, \rm repr})/F_{E, \rm hf}$ 
is the ratio of the observed fluxes at a given energy 
produced by the jet and reprocessing to that of the hot flow.

\end{enumerate}

\section{Summary}
\label{sec:summary}

A number of BH binaries have recently been found to show QPO features in their optical PSDs.
Recently, it has been shown that similar QPOs in the X-rays are well explained by
Lense-Thirring precession of the hot accretion flow.
Here we propose that OIR QPOs originate from the same process, namely that the outer parts 
of the precessing hot flow are radiating in the optical (by non-thermal synchrotron), 
producing QPOs at frequencies matching those in the X-rays.

We calculate the possible OIR profiles and make predictions for the phase difference with the X-rays.
The phase shift can be either 0 (at high inclinations) or $\pi$ (at low inclinations), with 
the boundary between these case depending on the azimuthal angle of the observer.
We also investigate the harmonic content and show dependence of the RMS on parameters for two first
harmonics. 
Here we show that the X-ray QPOs should have smaller RMS at the  fundamental frequency than the 
OIR QPOs. On the other hand, the X-ray QPOs should have much stronger harmonic.

We discuss possible QPO frequencies and their connection to the broadband spectral properties.
At the hard-to-soft state transition we expect the OIR QPO frequency to increase, always being
the same as in the X-rays, up to $\sim$10~Hz in the hard-intermediate state.
The broadband spectrum is expected to change in such a way that the emission at longer wavelengths
drops before the subsequent drop at shorter wavelengths (infrared, optical).
After that the X-ray transition occurs. 
In the hard state, the QPO feature is sometimes absent (not significant) in the X-ray PSD, but
detected in the optical.
Their common origin can be established through the cross-correlation analysis, where the
oscillations at corresponding frequency can be present.
Using future X-ray missions with high time-resolution capability, such as \textit{LOFT},
together with the corresponding high time-resolution instruments in the optical, we will be able to
confirm or discard this.

\acknowledgments

This work was supported by the Finnish Graduate School in Astronomy and Space Physics  (AV).
 

\end{document}